\documentclass[draft]{agujournal2019}
\usepackage{url} 
\usepackage{lineno}
\usepackage[inline]{trackchanges} 
\usepackage{soul}
\draftfalse

\journalname{JGR: Space Physics - Technical Reports: Methods}

\begin{document}

%
%

\title{A novel machine learning technique to identify and categorize plasma waves in spacecraft measurements}

\authors{Daniel Vech\affil{1} and David M. Malaspina\affil{1,2}}
\affiliation{1}{Laboratory for Atmospheric and Space Physics, University of Colorado Boulder, Boulder, CO, USA}
\affiliation{2}{Department of Astrophysical and Planetary Sciences, University of Colorado Boulder, Boulder, Colorado, USA}

\correspondingauthor{Daniel Vech}{daniel.vech@lasp.colorado.edu}







\begin{keypoints}
\item We develop a robust method to classify large data sets of power spectra of magnetic field
\item The technique significantly reduces the need for time consuming manual inspection of the data and allows the discovery of new wave forms
\item The classification technique can be used to categorize plasma waves based on frequency, amplitude and bandwidth
\end{keypoints}

%
%

%
%


\begin{abstract}
The available magnetic field data from the terrestrial magnetosphere, solar wind and planetary magnetospheres exceeds over $10^6$ hours. Identifying plasma waves in these large data sets is a time consuming and tedious process. In this Paper, we propose a solution to this problem. We demonstrate how Self-Organizing Maps can be used for rapid data reduction and identification of plasma waves in large data sets. We use 72,000 fluxgate and 110,000 search coil magnetic field power spectra from the Magnetospheric Multiscale Mission (MMS$_1$) and show how the Self-Organizing Map sorts the power spectra into groups based on their shape. Organizing the data in this way makes it very straightforward to identify power spectra with similar properties and therefore this technique greatly reduces the need for manual inspection of the data. We suggest that Self-Organizing Maps offer a time effective and robust technique, which can significantly accelerate the processing of magnetic field data and discovery of new wave forms.
\end{abstract}

\section*{Plain Language Summary}
[ enter your Plain Language Summary here or delete this section]

\section{Introduction}
Space missions like Wind \cite{lepping1995wind, bougeret1995waves}, Cassini \cite{dougherty2004cassini, gurnett2004cassini}, Van Allen Probes \cite{kletzing2013electric} and the Magnetospheric Multiscale Mission (MMS) \cite{russell2016magnetospheric, le2016search, lindqvist2016spin, ergun2016axial} collected massive data sets of electromagnetic field data in a very wide range of plasma environments. Given the sheer volume of the available data, identifying plasma waves is a difficult task since there is no straightforward method for automatic identification of wave activity such as fast mode, ion cyclotron, whistler, lower/upper hybrid and ion acoustic waves. In many cases the identification of plasma waves is still done by manually inspecting measurements of electric and magnetic fields. This is a time consuming and tedious process, which typically focuses only on a small subset of all the available data. Manual inspection of the data can certainly identify unique wave events (i.e. case studies) that display unusual features and provide useful insights into the underlying plasma physics. However, it is often difficult to design a search algorithm, which processes the entire available data set and filters out similar wave events to the ones, which were manually identified. 

Another frequently used approach for wave event selection is setting arbitrary thresholds on the data such as amplitude of magnetic and/or electric field fluctuations at given frequencies \cite<e.g.>[]{usanova2012themis}, magnetic compressibility \cite<e.g>[]{shi2017systematic,malaspina2017statistical,ala2019alfven}, magnetic helicity \cite<e.g.>[]{woodham2019parallel} and coherence between components of the electric and/or magnetic fields \cite<e.g.>[]{gao2014statistical}. Although, this technique is effective at processing large data sets, it may miss a significant number of waves because they may not meet the amplitude or other thresholds and the waves may occur at different frequency than expected.

In this Paper, we a present a solution to this problem: we use an unsupervised machine learning technique called Self-Organizing Map (SOM) \cite{kohonen1990self} and show how it can be used for rapid data reduction, identification of plasma waves from large data sets of spacecraft measurements and discovery of new wave forms. This technique sorts power spectra of magnetic field into nodes where power spectra belonging to the same node have similar properties: they all display wave activity at approximately the same frequency with similar amplitude, bandwidth and they have similar spectral indices and break frequencies. 

Various machine learning techniques have been developed for processing  solar radio bursts \cite{chen2017convolutional, mohd2020performance} and determining electron density from plasma waves \cite{zhelavskaya2016automated}, however, to the best of our knowledge, there has not been an attempt to process electromagnetic field data with SOMs. The most similar approach to the one presented in our paper was developed by the WAVES instrument team of the Wind spacecraft \cite{bougeret1995waves} who used an on-board neural network for the data selection of the time-domain sampler and thermal noise receiver. This neural network was trained to recognize event signatures, separate them from noise, rank them depending on how similar they are to the reference signal and mark them for download. Our technique is more advanced compared to this approach since there is no pre-defined reference signal and the data is organized into groups based on a similarity metric.

We demonstrate our approach by analyzing a data set of 72,000 fluxgate (FGM) and 110,000 search coil (SCM) magnetic field power spectra from the Magnetospheric Multiscale Mission (MMS$_1$). We compute the power spectra of the magnetic field for each 2-minute and 5-second interval for the FGM and SCM data, respectively and use the SOM to assign these power spectra to nodes. We show that this technique can classify power spectra remarkably well and makes it very easy to identify intervals with significant wave activity. Our technique is also suitable for classification of power spectra based on more subtle features such as break frequency and spectral indices.

\section{Self-Organizing Maps} \label{sec:floats}

In this Section we provide a summary of the basic principles of Self-Organizing Maps based on \citeA{kohonen2013essentials}. Self-Organizing Map is an unsupervised machine learning technique, which uses a finite set of models to represent a distribution of input data. A key feature of SOM is that the models are associated with the nodes of a regular (usually two-dimensional) grid. Adjacent nodes in the grid are associated with more similar models, in contrast, more distinct models are located farther away from each other in the grid. This is essentially a similarity diagram of the models, which allows us to understand the topographic relationships between high-dimensional input data. Each model is characterized by a set of weights that match the size of the input data items (i.e. for an input data set of 30 dimensional vectors, each model is characterized with 30 weights whose initial values are randomly chosen).

The central idea of the SOM learning process is the following: \emph{"Every input data item shall select the model that matches best with the input item, and this model, as well as a subset of its spatial neighbors in the grid, shall be modified for better matching."} The learning process is carried out with a batch-type method: the entire input data set is presented to the algorithm and all models and their weights are simultaneously updated. This process is repeated a few dozens times until the models are stabilized.

Various metrics are used to quantify the similarity between the input data and a model. In most applications including our study as well, the Euclidean distance ($d(q,p)=\sqrt{\sum(q_i - p_i})^2$) is used to measure the similarity. Therefore an input data item is assigned to a node, which has the lowest $d$ value (i.e. the input data item is most similar to that node). The determination of the number of nodes is carried out by a trial-and-error approach and a compromise must be made between resolution (number of nodes) and statistical accuracy (variability of input data assigned to the same node). For most applications the number of nodes is typically between a few dozens and a few hundreds. 

Potentially the most important advantage of SOMs is that the resulting grid clustering can be easily interpreted, therefore high dimensional input vectors with similar properties can be identified. A possible disadvantage is that SOMs require large volume of data in order to develop meaningful clusters. Additionally, the method assumes that the input data items fall in a finite number of classes and the different models can be made to correspond to these classes \cite{kohonen2013essentials}, however, this may not be the case if the input data is the affected by significant random noise.

\section{Demonstration of the Method}

We use 48000 burst mode intervals ($\approx$2400 hours of data) from MMS$_1$ spacecraft from 2015 October 1 to 2020 December 22. For the FGM (128 Hz cadence) data analysis, each burst mode interval is split into 2-min segments. If the data quality flag (assigned by the instrument team) is 0 (''good data'') for all data points in a 2-min segment, the magnetic power spectra of the X, Y, Z field components are calculated separately and are summed up hence the trace power spectra is obtained. For the calculation of the power spectra the Welch method (Welch, 1967) is used with Hanning window (50\% overlap). The FGM power spectra are calculated on a logarithmic grid with 83 frequencies in the range of 0.03 to 7 Hz.  For the SCM data (8192 Hz cadence), we use 5000 burst mode intervals ($\approx$155 hours of data) from MMS$_1$ spacecraft from 2015 October 1 to 2016 September 1. The SCM data processing follows the same methodology, however, the power spectra are calculated for each 5-second interval between 10-3000 Hz range. For both the FGM and SCM data we chose 83 frequency bins to ensure that fine structures in the power spectra (such as narrow band wave activity) are well preserved. We clip the FGM and SCM power spectra at 7 and 3000 Hz frequencies, respectively to avoid the classification of the spectra based on the frequency range, which is significantly affected by noise. For the data selection no constraints are imposed on the spacecraft location and both the FGM and SCM data sets include approximately 52\% and 48\%  power spectra from the dayside and nightside magnetosphere, respectively.

Since the background turbulence amplitude is highly variable, the power spectra have to be normalized before the SOM training process. We shift (in amplitude) the SCM and FGM power spectra with a constant factor so they are all set to log$_{10}$1 nT$^2$/Hz at the lowest frequency (0.03 and 10 Hz for the FGM and SCM, respectively). This normalization means that the differences between the power spectra are explained by differences in the spectral indices, break frequencies, wave activity and therefore the effect of the varying turbulence amplitude is minimized.

Our input data set includes 72,000 FGM and 110,000 SCM power spectra, respectively. For the SOM training process, we use 100 nodes, the Euclidean metric and 200 training iterations to ensure that the models are stabilized. We have tested several different number of iterations (50, 100, 150, 200) and found that after 150 iterations the models converged to a steady state and between 150 and 200 iterations there was negligible change in the distribution of the input data points among the nodes.

We have trained separate SOMs for the FGM and SCM data sets, respectively using all the available power spectra. The distribution of the input data points in the grid of nodes is shown in Figure 1. In Figure 1a and b the hexagonal grid shows the 100 nodes and the numbers correspond to how many power spectra were assigned to each node. The training process was repeated five times to test the robustness of the models. Approximately 0.08\% of the input vectors showed variations, meaning that they were not assigned to the same node in all iterations, therefore we suggest that the models are not sensitive to the initial values of the nodes.

We used the distribution of the input data in the SOM grid (Figure 1) to identify whether the number of nodes is sufficient. In the case of the FGM data, the nodes with the least number of data points have approximately $\approx$150 power spectra ($\approx$750 min of data). Given the size of the input data (72,000 power spectra), this means that rare wave events that make up only 0.2\% (150 / 72,000) of the data can form a separate node and therefore they can be easily identified. In the case of the SCM grid the ratio of the minimum number of power spectra / total number of power spectra is also around 0.2\%.

To understand how the SOM organized the input data among the 100 nodes, we calculate the average power spectra in each node. Figure 2 shows 9 examples of these and the error bars correspond to the standard deviation of the data in each frequency bin. As a reference, the dotted line shows the spectral index of -1.5. The spectral slopes are calculated over frequency windows, which have a factor of 2.8 and 1.8 width for the FGM and SCM data, respectively, which is the reason the beginning of the blue and brown lines in Figure 2 and 4 are not aligned.

In Figure 2a-b-c there are three nodes that show signatures of ion-scale wave activity, which can be seen an enhanced wave power in the frequency range of approximately 0.1-1 Hz and the spectra becomes flat. Considering the frequency range, these waves are consistent with ion cyclotron and fast mode waves \cite<e.g.>[]{gary1992mirror}. The second row of Figure 2 show three nodes, which have the most data points ($>$1500 power spectra in each node). These straight power spectra are typically associated with the dissipation of kinetic Alfv\'en waves and are often observed in the terrestrial magnetosheath (see Figure 4 in \cite{chen2017nature}). In the third row of Figure 2 there are three power spectra, which show steepening above 1 Hz. A possible explanation is ion cyclotron/fast mode wave activity in the frequency range of 0.1-1 Hz or in the absence of those waves the ion-scale break might be at an unusually high frequency.

Figure 3 shows a one-hour interval of fluxgate magnetic field data, the corresponding trace magnetic power spectra and the node number assigned to each 2-min interval. It can be see that the node number changes nearly every consecutive 2-min intervals, which indicates that the power spectra and therefore the fluctuations of the magnetic field show substantial variations. A major advantage of our technique is that this variability can be quantified and intervals with similar properties (i.e. same node number) can be qualitatively studied in terms of background plasma parameters (plasma beta, temperature anisotropy etc.) and spatial locations. Similar quantification of the power spectra with manual inspection is infeasible given the volume of the data. The threshold technique based on magnetic helicity \cite<e.g.>[]{woodham2019parallel}, compressibility \cite<e.g.>[]{malaspina2017statistical} and other proxies might be extended to an array of frequencies to select intervals based on multiple criteria (i.e measuring magnetic helicity simultaneously at 0.1, 1, 3 Hz etc.). However, as we use more and more variables to characterize the power spectra the interpretation is getting difficult as well. An important advantage of our technique is that despite the high-dimensional (83) input data, the SOM gives a clear interpretation by assigning each spectra to a node.

Figure 4 has the same format as Figure 2 and shows nine examples from the SCM nodes. We suggest that our technique is capable of distinguishing wave events based on amplitude, frequency, and bandwidth. Figure 4a-b-c show examples of nodes with minor wave activity. Figure 4d-e-f from left to the right show waves with increasing frequency of the spectral break and finally Figure 4g-h-i show wave events with increasing bandwidths. Designing a search algorithm, which takes into account the amplitude, frequency and bandwidth is a very challenging problem and would be a tedious trail-and-error process to fine tune the filtering parameters. In contrast, the SOM offers a simple solution to distinguish wave events based on these characteristics.

Our technique is capable of identifying rare wave events. For example, 95\% of the power spectra assigned to node $\#87$ (Figure 4i) occurred within a one hour interval out of the 155 hours searched. A key feature of SOM is that no prior assumptions are made about the underlying data. If certain wave events occur rarely in the data set but are significantly different from the rest of the data points, then they will be assigned to a separate node and therefore they can be easily identified. The search process based on thresholds would be inherently biased and would be based on assumptions about the expected wave spectral shapes in the data set, therefore it would be difficult to identify rare events.

Our technique is also suitable for identifying intervals with desirable frequency features. For example, we identified intervals when the power spectra moved from node $\#42$ (Figure 5a) to node $\#41$ (Figure 5b). Figure 5c shows the time-frequency domain for these intervals: the first 5s (node $\#42$) shows signatures of wave activity starting from 30 Hz and between 5-10s (node $\#41$) the wave activity increases toward higher frequencies and reaches nearly 200 Hz. For the design of a search algorithm several free parameters (e.g. lowest frequency of the wave, rate of change of the frequency) would have to be estimated with a reasonable accuracy to identify large number of similar wave events. In the case of SOM analysis, there is no need to estimate these parameters for the purpose of wave identification since one can look for specific frequency features given the nodes.

\begin{figure}
    \centering\includegraphics[width=0.9\linewidth]{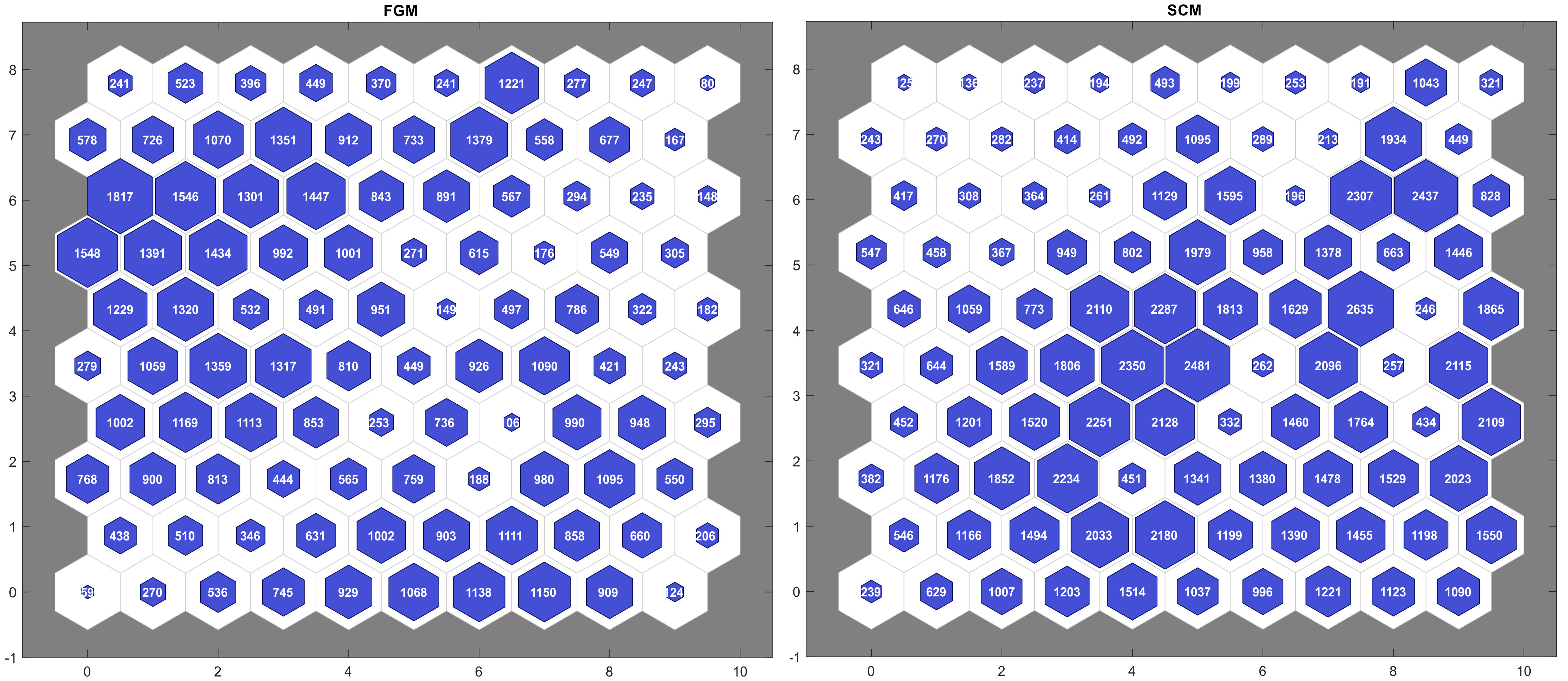}
\caption{a)-b) Distribution of the input data in the grid of the FGM and SCM self-organized maps. The 100 nodes are numbered from left to right the following way: bottom left corner is node $\#1$, bottom right corner is $\#10$. Top left and top right nodes are $\#91$ and $\#100$, respectively.}
  \label{fig:1}
\end{figure}

\begin{figure}
    \centering\includegraphics[width=0.9\linewidth]{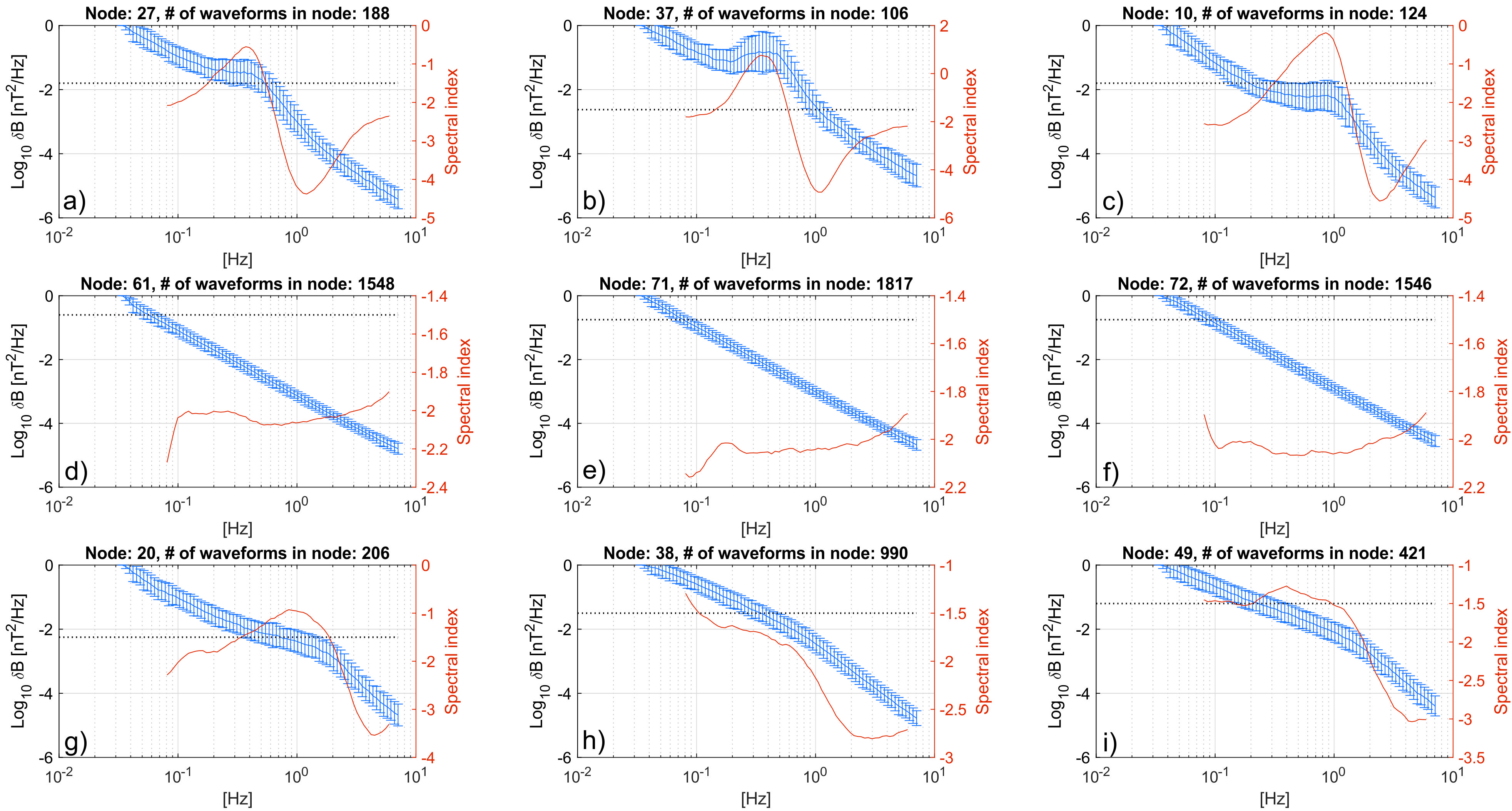}
\caption{Averaged FGM power spectra and corresponding error bars from nine nodes.}
  \label{fig:2}
\end{figure}

\begin{figure}
    \centering\includegraphics[width=0.9\linewidth]{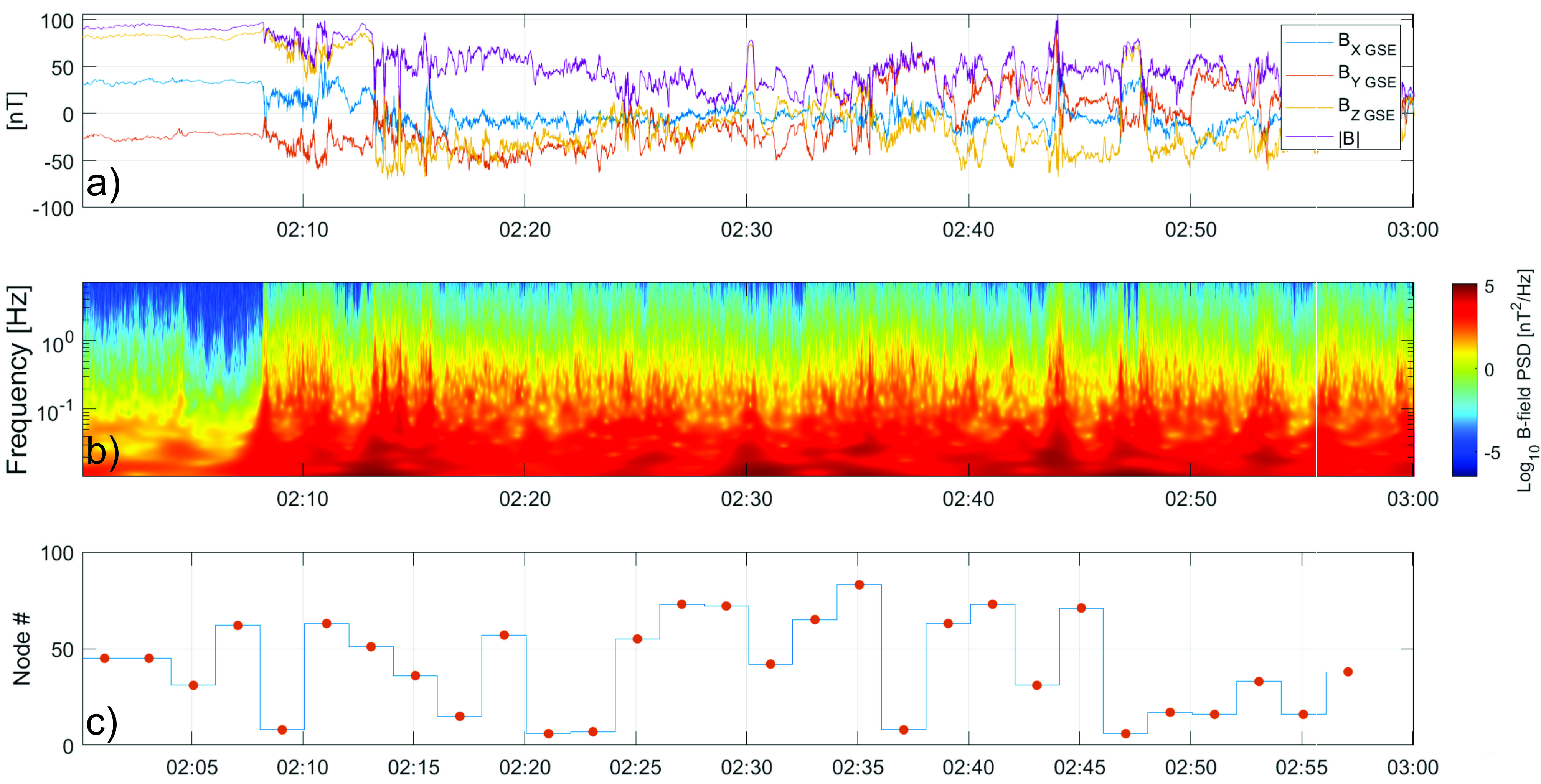}
\caption{a) Example of a one-hour long fluxgate magnetic field data from 10 November 2015 02:00-03:00, (b) the corresponding trace power spectra and (c) the assigned node numbers to each 2-min interval. In panel c) the red circles are placed at the center of each 2-min interval.}
  \label{fig:3}
\end{figure}

\begin{figure}
    \centering\includegraphics[width=0.9\linewidth]{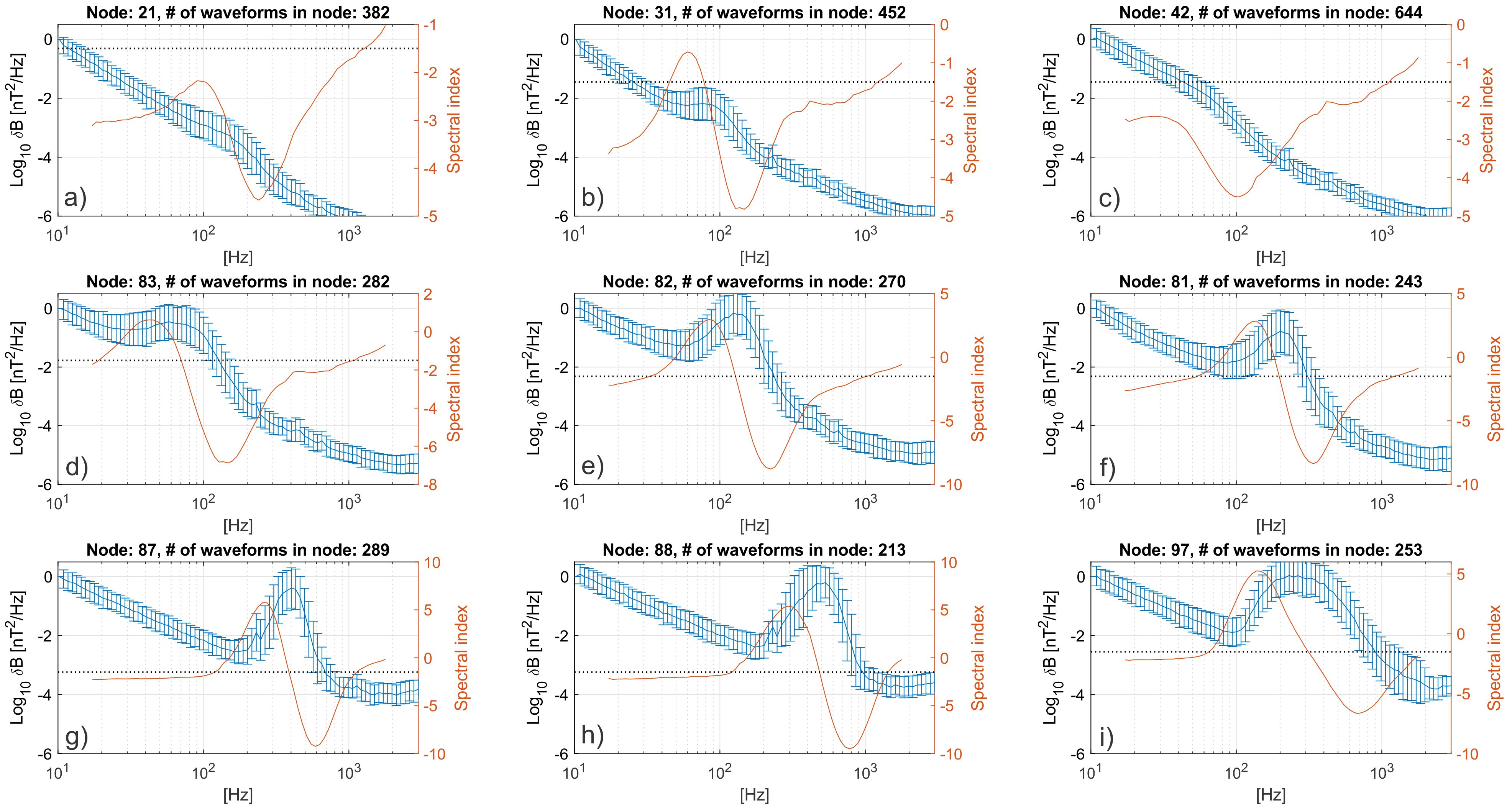}
\caption{Averaged SCM power spectra and corresponding error bars from nine nodes.}
  \label{fig:4}
\end{figure}

\begin{figure}
    \centering\includegraphics[width=0.9\linewidth]{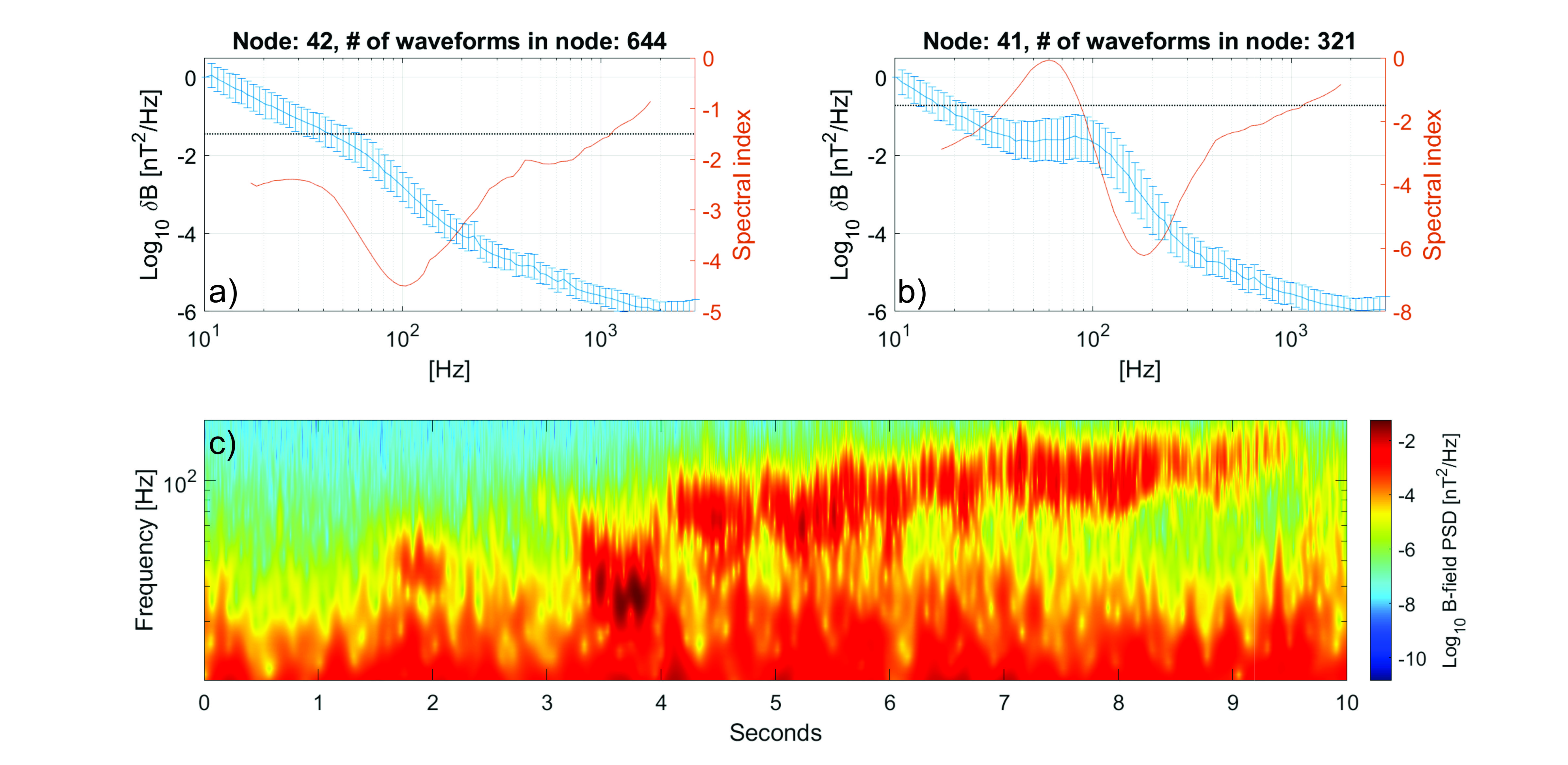}
\caption{Example for a wave with time dependent frequency feature based on SCM data. The first 5s of the interval was assigned to node $\#$42 while the consecutive interval to $\#$41. In the first interval the wave activity starts at approximately 30 Hz and steadily increases toward 200 Hz in the second interval.}
  \label{fig:5}
\end{figure}

\section{Conclusion} \label{sec:displaymath}

In this Paper, we have described a novel machine learning technique to classify the power spectra of magnetic field fluctuations. The technique uses a high-dimensional input (magnetic turbulence amplitude in 83 frequency bins) and organizes the power spectra into 100 nodes based on their shape. Similar classification of the data is not possible with the threshold techniques or with manual inspection of the data. We note that the technique does not identify the wave mode itself, which would require careful analysis of the wave polarization properties and dispersion relations. Despite this limitation, the proposed technique can have huge implications to process magnetic field measurements by significantly reducing the need for time consuming manual inspection of the data and eliminating the need for arbitrary thresholds on the data to filter out wave events. Here we summarize three main areas where our technique will advance the processing of electric and magnetic field data.

Identification of wave type: The proposed technique can greatly accelerate the processing of electromagnetic field data by identifying and classifying consistently occurring patterns in plasma wave activity independent from the observational biases or expectations of a given researcher. In our example, 3 nodes of the FGM data showed significantly enhanced wave power and nearly flat power spectra in the 0.1-1 Hz frequency range, which are consistent with ion cyclotron and fast magnetosnic waves. For the SCM data we found large number of nodes with intense wave activity and we showed that our technique can classify them based on frequency, bandwidth, and normalized amplitude as well.

Discovery of new wave spectral shapes: a very important feature of the proposed approach is that no assumptions are made about the shape of the power spectra, which is a major advantage compared to the threshold techniques. Regardless the underlying input data, the SOM automatically identifies wave events that deviate significantly from the rest of the input data and assigns them to nodes accordingly. This means that if the data set includes only a handful of those wave events we are looking for, they can still be identified. This aspect is particularly useful for studying less well understood space plasmas such as the inner-heliosphere \cite{bale2016fields} where the observable wave forms might be substantially different from the expectations and using arbitrary threshold could not identify those waves events. The technique could lead to an improved understanding of magnetospheric physics as well since most studies rely on arbitrary thresholds of frequency, amplitude, ellipticity for wave identification \cite<e.g.>[]{usanova2012themis, shi2017systematic} and therefore compelling wave events are hidden in the large data sets provided by Van Allen Probes, Cluster, THEMIS and MMS.

Our technique is capable of classifying power spectra based on spectral indices and break frequencies. Very large number of studies focused on the ion-scale spectral break and its dependence on plasma parameters \cite<e.g>[]{bourouaine2012spectral, vech2018magnetic, duan2020radial}. One fundamental difficulty is that measuring the break frequency is challenging especially in those cases when the spectral index shows only minor steepening around ion scale. Our technique makes it possible to identify groups of power spectra with similar break frequencies therefore the underlying plasma properties can be investigated and quantified. Additionally, studying the underlying plasma parameters for each of the 100 nodes can help to understand the operation of the dissipation range turbulence cascade and damping mechanisms since we could precisely quantify how the power spectra responds to changes in plasma beta, ion and electron temperature anisotropy and other parameters.


\acknowledgments
D. V. was supported by NASA contract 80NSSC21K0454. D.M. was supported by NASA contract 80NSSC19K0305. The data presented in this paper are the L2 data of MMS, which can be accessed from https://lasp.colorado.edu/mms/sdc/public/)


%
%


%
%
%
%
%

\end{document}